\documentclass{article}
\usepackage{times}
\usepackage{soul}
\usepackage{url}
\usepackage[hidelinks]{hyperref}
\usepackage[utf8]{inputenc}
\usepackage[small]{caption}
\usepackage{graphicx}
\usepackage{amsmath}
\usepackage{amsthm}
\usepackage{booktabs}
\usepackage{algorithm}
\usepackage{algorithmic}
\usepackage[capitalize]{cleveref}
\usepackage{amssymb, mathtools}
\usepackage{etoolbox}
\usepackage{xcolor}
\usepackage{comment}
\usepackage{enumerate}
\usepackage{multicol}
\usepackage[normalem]{ulem}
\usepackage{xspace}
\usepackage{bbm}
\usepackage{fdsymbol}
\usepackage{yfonts}

\theoremstyle{plain}
\newtheorem{theorem}{Theorem}
\newtheorem{corollary}[theorem]{Corollary}
\newtheorem{lemma}[theorem]{Lemma}
\newtheorem{fact}[theorem]{Fact}

\newtheorem{observation}[theorem]{Observation}

\theoremstyle{definition}

\theoremstyle{remark}

\newcounter{rscounter}
\newcounter{dbcounter}
\newcommand{\ontology}{\database, \ruleset}

\definecolor{salmon}{HTML}{b35470}

\newcommand{\childrenamount}{2}
\newcommand{\childrendots}{
	\ifnum \childrenamount=2
	,
	\else
	,\dots,
	\fi
}

\newcommand{\DLCL}{DL-Lite$_{\text{core}}\;$}

\title{On the entailment problem for DL-Lite$_{core}$ ontologies and conjunctive queries with negation.}

\author{%
Jerzy Marcinkowski\\ \text{\small jma@cs.uni.wroc.pl}\\ University of Wrocław\and
Piotr Ostropolski-Nalewaja \\
\text{\small  postropolski@cs.uni.wroc.pl} \\ University of Wrocław
}

\begin{document}

\maketitle

\begin{abstract}
We show that the entailment problem,  for a given \DLCL ontology, and given conjunctive query with inequalities, is undecidable.

We also show that this problem remains undecidable if conjunctive queries with safe negation are considered instead of
conjunctive queries with inequalities.
\end{abstract}

\newcommand{\alfab}{{\textfrak A}}

\newcommand{\thue}{{\mathbb T}}

\newcommand{\szymeq}{{\;\simeq_\Pi\;}}

\newcommand{\krzero}{{\mathcal KB}_0 }

\newcommand{\ddd}{\mathbbmss{d}}
\newcommand{\nnn}{\mathbbmss{n}}
\newcommand{\mmm}{\mathbbmss{m}}
\newcommand{\ccc}{\mathbbmss{c}}
\newcommand{\kkk}{\mathbbmss{k}}
\newcommand{\jjj}{\mathbbmss{j}}
\newcommand{\ddelta}{\bar{\delta}}

\newcommand{\DDD}{\mathbb D}
\newcommand{\VVV}{\mathbb V}

\newcommand{\strz}[3]{#1 \stackrel{#2}{\longrightarrow} #3 }

\renewcommand{\ontology}{${\mathcal O}^{\neq} $}
\newcommand{\ontologyy}{${\mathcal O}^{\neg} $}

\newcommand{\slott}{{\mathbb S}}

\newcommand{\iksA}{x_{\footnotesize \text{\scalebox{1}{$\diamondsuit$}}}}

\newcommand{\jest}[1]{#1^{\text{\scalebox{0.8}{$\exists$}}}}
\newcommand{\ontol}{{\mathcal O}}

\newcommand{\kontologia}{{\mathcal K}}
\renewcommand{\diamondsuit}{{\textcolor{gray}{\vardiamondsuit}}}

%

\section{Our contribution.}



We prove the following four theorems:

\begin{theorem}\label{main1}
The problem:

\vspace{0.6mm}
\noindent
{\em Given an \DLCL ontology \ontology \;and a Boolean conjunctive query with inequalities $\psi$. Does \ontology$\;\models \psi$?
}

\vspace{0.6mm}
\noindent
is undecidable.
\end{theorem}

\begin{theorem}\label{main2}
The problem:

\vspace{0.6mm}
\noindent
{\em Given an \DLCL ontology \ontologyy \;and a Boolean conjunctive query with safe negation $\phi$. Does \ontologyy$\models \phi$?
}

\vspace{0.6mm}
\noindent
is undecidable.
\end{theorem}

\begin{theorem}\label{niemain1}
The problem:

\vspace{0.6mm}
\noindent
{\em Given an \DLCL ontology $\ontol$ and a  formula $\Psi$ which is a union of Boolean conjunctive queries with inequalities. Does $\ontol \models \Psi$?}

\vspace{0.6mm}
\noindent
is undecidable.
\end{theorem}

\begin{theorem}\label{niemain2}
The problem:

\vspace{0.6mm}
\noindent
{\em Given an \DLCL ontology $\ontol$ and a  formula $\Phi$ which is a union of Boolean conjunctive queries with safe negation. Does $\ontol \models \Phi$?}

\vspace{0.6mm}
\noindent
is undecidable.
\end{theorem}

One might think that Theorems \ref{niemain1} and \ref{niemain2} are redundant, as every CQ is obviously a UCQ.
But there is a certain nuance here, which will now be discussed.

\subsection{Certain nuance regarding the assumptions behind Theorems \ref{main1} and \ref{main2}.}\label{nuance}


For an ontology
$\kontologia$ and a query
 $\xi$, we say that $\kontologia$ entails $\xi$, denoted as
$\kontologia\models \xi$,
 if for every structure $D$, being a model of $\kontologia$ (also denoted as
 $\kontologia\models D$), query $\xi$ is true in $D$ (for some reason, also denoted as  $ D\models \xi$).

 But what does it mean that $D$ is a model of $\kontologia$?
Suppose $\kontologia$ contains the assertion:
 \begin{center}
 $Knows(dr\_jekyll, mr\_hyde)$
 \end{center}
  for some binary
 relational symbol $Knows$ and for two constants, $dr\_jekyll$ and $ mr\_hyde$. Consider
 a structure $D$, containing just one vertex, which is an interpretation of both
the constants, together with the single
 atom of the relation $Knows$. Is $D$ a model of $\kontologia$?

 If we accept that different constants mentioned in the ontology can
be interpreted as the same element of
 the structure, then no ontology can ever entail any conjunctive query
with inequality. This is
 because  the ``well of positivity'', that is a single vertex, for
which all atomic formulae are true,
 is always a model of such ontology, and of course no conjunctive
query with inequality can be true in such structure.

 For this reason, for Theorem 1 to have any chance to be true, we need to adopt the
Unique Name Assumption, which exactly says that
 different constants must be interpreted as different vertices of the
model\footnote{Unique Name Assumption is, as we understand, a standard
choice in the context of the logic \DLCL.}.

 On the other hand, Theorem \ref{niemain1} holds regardless of whether we adopt the
Unique Name Assumption, and our proof remains valid in either case. The idea here is that the UCQ constructed by
our reduction is a disjunction of some conjunctive queries with inequalities and of some conjunctive queries without inequalities. And then, the more
someone tries to identify the verticies of some structure, the more false the disjuncts with inequalities get, but, at the same time,
the more true the ones without inequality become.

Regarding Theorem \ref{main2} we face a very similar phenomenon.
Suppose $\kontologia$ contains an assertion:
\begin{center}
$Husband(Oedipus,  Jocasta)$
\end{center}

but does not contain the assertion:

\begin{center}
$\hfill Son(Oedipus,  Jocasta)$ \hfill (*)
\end{center}

Will we accept, as a model of $\kontologia$ a structure, with two
constants, $ Oedipus$ and
 $ Jocasta$, in which
(*) is true? If we agree that ground facts, which involve only
constants mentioned in the ontology, but which are not present in the
ontology, may be true in the model, then no ontology can ever entail
any conjunctive query with safe negation. This is because there is
no mechanism in \DLCL able to prevent the structure from containing
all the facts involving elements of its active domain, and of course
in such structure no conjunctive query with safe negation can be true.

For this reason, for Theorem \ref{main2} to be true, we need to adopt the Partial Closed World Assumption, which says that no new facts
involving only constants present in the $\kontologia$ may be true in the model\footnote{We understand it is not a standard \DLCL  assumption.} But of course, under this assumption, the model can contain new facts, provided that they also involve elements
which are not interpretations of the constants from $\kontologia$.

On the other hand, Theorem \ref{niemain2} holds regardless of whether we adopt the Partial Closed World Assumption or not.

 \subsection{How the rest of the paper is organized}

 Sections \ref{sec:source} - \ref{secmain2} are devoted to the proofs of Theorems \ref{main1} - \ref{niemain2}. First, in Section \ref{sec:source}
 we present the undecidable problem we use for the reduction, the word problem for finitely generated semigroups.

In Section \ref{systemy} we prepare our toolbox. And, at the same time, we  almost effortlessly prove Theorems \ref{niemain1} and \ref{niemain2}.

Then we proceed with Section \ref{secmain1}, where Theorem \ref{main1} is proven and Section
 \ref{secmain2} where proof of Theorem \ref{main2} is presented.

Finally, in Section \ref{sec:finite}, we discuss the finite entailment variant.


\section{The source of undecidability}\label{sec:source}

Our source of undecidability will be the word problem for finitely generated semigroups, also known as the word 
problem for Thue systems. 

An instance of this problem is a triple $\thue= [l,r, \Pi]  $, where $l,r\in \alfab^*$ are nonempty words over some
finite alphabet $\alfab$, and where $\Pi =\{[l_1,r_1], [l_2,r_2],\ldots [l_\kkk,r_\kkk]\}$ is a set of pairs
of nonempty words, where again  $l_k,r_k\in \alfab^*$  for each $1\leq k \leq \kkk$.

For a set $\Pi$ as above, the relation $\szymeq$ is defined as the smallest eqivalence relation such that,
for each $w,v\in \alfab^*$ and for each $1\leq k \leq  \kkk$ it holds that $ wl_kv \szymeq wr_kv $.

An instance  $\thue= [l,r, \Pi]  $ is {\em positive} when $l \szymeq r$ and is {\em negative} otherwise.

It is a well-known fact that:

\begin{fact}\label{fact}
 For a set $\Pi$ as above, and for $w,w'\in  \alfab^*$ the relation $w\szymeq w'$ holds if and only if $w'$ can be produced from $w$ in a finite number of rewriting steps,
where in each step $vl_kv'$ gets rewritten into  $vr_kv'$, for some $v,v'\in \alfab^*$ and for some $1\leq k\leq \kkk$, or the other way round, that is $vr_kv'$ gets rewritten into  $vl_kv'$.
\end{fact}

It is also well known that the problem, whether a given instance of the word
problem for Thue systems is positive, is undecidable. See for example \cite{handbook}.

\vspace{1.1mm}
\noindent
{\bf  From now on  an instance  $\thue= [l,r, \Pi]  $ is considered fixed}. This means that also $\alfab$ is fixed,
and we denote $\mmm= |\alfab|$.

\vspace{1.1mm}
In order to prove Theorem \ref{main1}  we will construct, in Section \ref{secmain1}
a \DLCL ontology \ontology and a Boolean conjunctive query with inequalities $\psi$ such that:

\begin{center}
$\thue$ is a positive instance ~~~ if and only if ~~~ \ontology$\models \psi$.
\end{center}

In order to prove Theorem \ref{main2}  we will construct, in Section \ref{secmain2}
a \DLCL ontology \ontologyy and a Boolean conjunctive query with safe negations $\phi$ such that:

\begin{center}
$\thue$ is a positive instance ~~~ if and only if ~~~ \ontologyy$\models \phi$.
\end{center}

\section{Building the tools}\label{systemy}

\subsection{Candidate structures and perfect structures}\label{sec:candidate}

Define the relational signature $\bar\alfab = \alfab\cup \{A, T\}$, where
the relation symbols from the set $\alfab$ are binary, the relation symbol $A$ is unary,
and the relation symbol $T\not\in\alfab$ is binary.  $T$ will play no role outside of Section \ref{secmain2}, so
you can safely ignore it for some time.  The fact that we are
using the same set $\alfab$ here as in Section \ref{sec:source} is not a mistake: we really want the names of the relations to
coincide with the names of the symbols from the alphabet $\alfab$ of the Thue system.

By a {\em candidate structure }
we will mean a relational structure  $D$, over $\bar\alfab$, with the set of vertices $V$,
for which the following
conditions are satisfied:

\vspace{1mm}
\noindent
(p1)   $D\models A(a) $ and  $D\models T(a,a)$ for some $a\in V$;

\vspace{1mm}
\noindent
(p2) For each $s\in V$ and for each $S\in \alfab$ if $D\models A(s)$
then there exists $t\in V$ such that $D\models S(s,t)$;

\vspace{1mm}
\noindent
(p3) For each $s,t\in V$ and  each $R,S\in \alfab$ if $D\models R(s,t)$
then there exists $u\in V$ such that $D\models S(t,u)$;

\vspace{2mm}
A careful reader may have noticed at this point that the property of being a candidate structure is expressible in \DLCL. Indeed, denote as $\mathcal O$ the \DLCL ontology comprising the rules:

\vspace{2mm}
\noindent
\textbullet~
$A(a)$ and $T(a,a)$

\vspace{1mm}
\noindent
\textbullet~
$A\sqsubseteq \exists R$ \hfill for each relation symbol $R\in \alfab$

\vspace{1mm}
\noindent
\textbullet~
$\exists S^{-1} \sqsubseteq \exists R$ \hfill for each relation symbols $S,R\in \alfab$

\vspace{2mm}
Then $D\models \mathcal O$ if and only if $D$ is  a candidate structure.

\vspace{2mm}
For two vertices $s,t$ of some structure $D$ over $\alfab$ and for $R\in \alfab$  we will write $\strz{s}{R}{t}$ if
$D\models R(s,t)$. For two vertices $s,t$ of some structure $D$ over $\alfab$, for $w\in \alfab^*$,  and for $R\in \alfab$,  we will write $\strz{s}{wR}{t}$ if
there exists $u$ such that $\strz{s}{w}{u}$ and $\strz{u}{R}{t}$.

Notice that for each $w\in \alfab^*$ the expression
$\strz{x}{w}{y}$ is a conjunctive query (without inequalities and without negation) with free variables $x$ and $y$.

It easily follows from conditions (p2) and (p3) above that:

\begin{observation} \label{obs:dodana}
For each candidate structure $D$, with the set $V$ of vertices, for each words $w,u\in  \alfab^*$ and each
$s\in V$ such that  $\strz{a}{w}{s}$,
there
exists at least one vertex $t$ of $D$ such that  $\strz{s}{u}{t}$.
\end{observation}

A candidate structure $D$, with the set $V$ of vertices, will be called a {\em perfect structure} if for each $w\in \alfab^*$,
 each $s,t\in V$ such that $\strz{a}{w}{s}$ and each
$1\leq k \leq \kkk$:

 $\;\;\;\strz{s}{l_k}{t}\;\;\;$ if and only if  $\;\;\;\strz{s}{r_k}{t}$

Notice that perfection is entirely in the eyes of the vertices reachable from $a$.

A candidate structure which is not perfect will (naturally) be called {\em imperfect}.

Now, the following lemma is not going to surprise anyone:

\begin{lemma}\label{lem:perfection} Suppose $D$ is a perfect structure with the set of vertices $V$.
Suppose $s\in V$ and $w,v\in \alfab^*$ are such that
 $\strz{a}{w}{s}$ and $w \szymeq v $. Then  $\strz{a}{v}{s}$.
\end{lemma}

\noindent {\sc Proof:} Use Fact \ref{fact} and proceed by induction with respect to the number of rewriting steps needed to
produce $v$ from $w$.\qed

 \subsection{A remark about conjunctive queries with one free variable. }\label{sec:remark}

 \vspace{1mm}
 For a conjunctive query $\xi(x)$, possibly with inequalities or safe negation,
 with one free variable,
 we denote by $\jest{\xi}$ the Boolean conjunctive query $\exists x\;\xi(x)$.

 Variable $x$ will
 be called {\em the distinguished variable} in $\jest{\xi}$.

Now imagine we have two such queries  $\xi_1(x_1)$ and  $\xi_2(x_2)$, each with one free variable.

And consider a new boolean conjunctive query:
$$ \xi \;\;\; = \; \;\; \exists x_1,x_2\;\;\; x_1\neq x_2 \;\; \wedge \;\; \xi_1(x_1) \;\;  \wedge  \;\; \xi_2(x_2) $$
Then (and this may seem to be an obvious observation, but will be very useful in
Sections \ref{secmain1} and \ref{secmain2}),
in order to satisfy query $\xi$, in some structure $D$,
two homomorphisms $h_1: \jest{\xi_1} \rightarrow D $ and $h_2: \jest{\xi_2} \rightarrow D$ need to be found,
such that $h_1(x_1)\neq h_2(x_2)$. Apart from the last inequality, which concerns the
 distinguished variables of the two queries,
the homomorphisms are independent,
because even if (syntactically)  $\jest{\xi_1}$ and $\jest{\xi_2}$  shared some variables, the shared variables
are local in $\xi_1(x_1)$ and in  $\xi_2(x_2)$.

In such context, we will call queries $\jest{\xi_1}$ and $\jest{\xi_2}$ {\em components} of $ \xi$.
Of course a conjunctive query may have more than two components.

\subsection{$\diamondsuit$-structures}

A candidate structure $D$, with the set of vertices $V$, will be called a {\em $\diamondsuit$-structure} if there
exist $s,t\in V$ such that:
$$D\;\models \;\; A(s) \;  \wedge \; \strz{s}{l}{t} \; \wedge \; \strz{s}{r}{t}$$
Being a $\diamondsuit$-structure is of course expressible by a Boolean conjunctive query. This query will be so important for us
that,  for sake of future notational simplicity we will give it two names\footnote{Apparently, the Unique Name Convention does not apply to our formulas.},
$\jest{\gamma_\diamondsuit}$ and $\jest{\beta_\diamondsuit}$   where:
$$ \gamma_{\diamondsuit}(\iksA)  = \beta_{\diamondsuit}(\iksA) = \exists  y \; A(\iksA)\;\wedge\; \strz{\iksA}{l}{y}\; \wedge \; \strz{\iksA}{r}{y} $$

\vspace{1mm}
As an obvious consequence of Lemma \ref{lem:perfection} we get:

\begin{observation} \label{lem-cztery}{observation}
If  $\thue$ is positive and $D$ is a perfect structure then there exists a vertex $s$ of $D$ such that
$\strz{a}{l}{s}$ and $\strz{a}{r}{s}$.
\end{observation}

The following is a direct corollary:

\begin{corollary} \label{lem-cztery-cor}
In other words, if  \;$\thue$ is positive and $D$ is a perfect structure, then $D\models \jest{\gamma_\diamondsuit}$.
\end{corollary}

What if $\thue$ is a negative instance? This is what the next subsection is about.


\subsection{The canonical structure $\DDD$.}

We will now define a  structure $\DDD$  over $\bar{\alfab} $ , with set of vertices $\VVV$, which is ``canonical'' for $\thue$.
Take $\VVV= \alfab^*/{\szymeq}$, that is the set of all equivalence classes of the relation  $\szymeq$.

Now, once we know what the vertices of $\DDD$ are, let us define its edges.  For each
$w,v \in \alfab^* $ and each $R\in \alfab$ define:
$$\DDD\models R([w]_{\szymeq}, [v]_{\szymeq})  \text{\;\;\;if
and only if\;\;\;} wR \szymeq v$$

One needs to notice here that the above definition
is independent of the choice of representatives $w$ and $v$, from their respective  equivalence classes.
So assume that  $wR \szymeq v$ and take $w',v'$ such that $w\;\szymeq \; w'$ and $v\;\szymeq \; v'$. Then:
$$w'R\;\; \szymeq \;\;  wR \;\; \szymeq \;\; v \;\; \szymeq \;\; v'$$

\noindent
Define also:

$$\DDD\models A([w]_{\szymeq})  \text{\;\;\;if
and only if\;\;\;} w \szymeq \varepsilon$$

\noindent
where $\varepsilon$ denotes the empty word.
Finally, define:

$$\DDD\models T(a,s)\;\; \text{ for each} \;\; s \in \VVV$$

It now easily follows from the construction that:

\begin{lemma} \label{lem-dwa}
(i) $\DDD$ is a perfect structure;

\noindent
(ii) if $\thue$ is a negative instance then $\DDD \not\models \jest{\gamma_{\footnotesize \diamondsuit }} $.
\end{lemma}


\subsection{Detecting imperfections, part 1. Proof of Theorem \ref{niemain1}.} \label{sec:detecting1}

\vspace{1mm}
For $1\leq k \leq \kkk$ define $ \gamma_k $ as the conjunctive query with inequality, with one free variable $x_k$:
$$ \gamma_k(x_k) \;\;\;\; = \;\;\;\; \exists y,y' \;\;\; y\neq y' \; \wedge \; \strz{x_k}{l_k}{y} \; \wedge \; \strz{x_k}{r_k}{y'} $$

Finally, let $\Gamma^{\footnotesize \neq}$ be the UCQ with inequalities:

$$\Gamma^{\neq}  \;\;\; = \;\;\; \bigvee_{1\leq k \leq \kkk}\; \jest{\gamma_k} $$

It now follows easily from Observation \ref{obs:dodana} and the definition of perfect structure:

\begin{observation}\label{obs:brakuje1}
 If a candidate structure $D$, with $V$ being its set of vertices, is imperfect, then there exist
 $w\in \alfab^*$,
 $s\in V$,
 and $1\leq k \leq \kkk$
 such that $\strz{a}{w}{s} $ and
 $D\models  \gamma_{k}(s)   $.
 \end{observation}

Observation \ref{obs:brakuje1} implies:


\begin{observation} \label{obs-trzy}
For a candidate structure $D$, if  $D$ is imperfect then $D\models \Gamma^{\neq} $.
\end{observation}

The opposite implication does not hold, it is easy to imagine a perfect structure in which
$\Gamma^{\neq}$ is satisfied. But it is easy to see that:

\begin{observation} \label{obs-jeden}
$\DDD \not\models \Gamma^{\neq}$.
\end{observation}

\vspace{1mm}
Let now $\;\;\Psi \;\; = \;\; \Gamma^{\neq} \;\vee\; \jest{\gamma_\diamondsuit} $. ~~~Then:
\vspace{1mm}

\begin{lemma}\label{koniec3}
${\mathcal O}\models \Psi $ if and only if $\thue$ is a positive instance.
\end{lemma}

\noindent
{\sc Proof:} First assume that $\thue$ is a negative instance. Then $\DDD \not\models \Psi$. This is because $\DDD \not\models \Gamma^{\neq}$ (from Observation
\ref{obs-jeden}) and $\DDD \not\models \jest{\gamma_\diamondsuit} $  (from Lemma \ref{lem-dwa}).

Now assume that instance $\thue$  positive, and take any candidate structure $D$. If $D$ is imperfect then $D\models \Gamma^{\neq} $ (from
Observation \ref{obs-trzy}) and hence $D\models \Psi$.  If $D$ is perfect then $D\models \jest{\gamma_\diamondsuit}$  (from Corollary \ref{lem-cztery-cor})
and hence $D\models \Psi$. \qed

\vspace{2mm}
Notice that {\bf we have}  just {\bf proved Theorem \ref{niemain1}}.
Unsurprisingly, whenever $\thue$ is an negative instance we can use $\DDD$ as a \emph{countermodel} for entailment. In fact, the negative direction (Lemmas \ref{lem:niemain2}, \ref{lem:main1-neg}, and \ref{lem:main2-neg}) is easier than the positive, because, as it turns out, we could always use $\DDD$ as a countermodel.
Also notice, that it did not matter in the above proof whether we adopt the
Unique Name Assumption or not. 


\subsection{Detecting imperfections, part 2. Proof of Theorem \ref{niemain2}.}\label{sec:detecting2}

Suppose $l'_k, r'_k\in \alfab^*$ and $L_k, R_k\in \alfab$ are,  for each $1\leq k \leq \kkk$, such that  $l_k = l'_kL_k $ and  $r_k = r'_kR_k $. Recall that $l_k$ and $r_k$ are nonempty, so we can always split them in this way.

For $1\leq k \leq \kkk$ define $ \beta_{[r,k]}(x_{[r,k]}) $ as the conjunctive query with safe negation, with one free variable $x_{[r,k]}$:
$$ \beta_{[r,k]}(x_{[r,k]}) \;= \; \exists y,y' \; \strz{x_{[r,k]}}{l_i}{y} \; \wedge \; \strz{x_{[r,k]}}{r'_i}{y'} \; \wedge \; \neg R_k(y',y) $$
Analogously, for $1\leq k \leq \kkk$ define $ \beta_{[l,k]}(x_{[l,k]}) $ as the conjunctive query with  safe negation, with one free variable $x_{[l,k]}$:
$$ \beta_{[l,k]}(x_{[l,k]}) \;= \; \exists y,y'\;  \strz{x_{[l,k]}}{l'_i}{y} \; \wedge \; \strz{x_{[l,k]}}{r_i}{y'} \; \wedge \; \neg L_k(y,y') $$
Finally, let $\Gamma^\neg$ be the UCQ with inequalities:

$$\Gamma^\neg  \;\;\; = \;\;\; \bigvee_{1\leq k \leq \kkk}\; \jest{\beta_{[l,k]}} \vee  \jest{\beta_{[r,k]}} $$

It now follows easily from Observation \ref{obs:dodana} and the definition of perfect structure:

\begin{observation}\label{obs:brakuje2}
 If a candidate structure $D$, with $V$ being its set of vertices, is imperfect, then there exist
 $w\in \alfab^*$,
 $s\in V$,
 and $1\leq k \leq \kkk$
 such that $\strz{a}{w}{s} $ and:
 $$D\models  \beta_{[l,k]}(s) \vee  \beta_{[r,k]}(s)  $$
 \end{observation}

Observation \ref{obs:brakuje2} implies:

\begin{observation}
For a candidate structure $D$, if $D$ is imperfect then  $D\models \Gamma^\neg $.
\end{observation}

Again, the opposite implication obviously does not hold.

\begin{observation}\label{obs:nienazwana}
$\DDD \not\models \Gamma^\neg$
\end{observation}

Let now $\Phi \;\; = \;\; \Gamma^\neg \vee \jest{\beta_\diamondsuit} $. Then the proof of the following Lemma is analogous
to the proof of Lemma \ref{koniec3}:

\begin{lemma}\label{lem:niemain2}
${\mathcal O}\models \Phi $ if and only if $\thue$ is a positive instance.
\end{lemma}

Notice that {\bf we have} just {\bf proved Theorem \ref{niemain2}}.
Notice also that again it did not matter in the above proof, whether we adopt the
Partial Closed World Assumption or not.


\subsection{Structures $\slott_n$ and $\DDD_{\nnn}$. And ontology $\Omega^{\nnn}$.}\label{sec:slots}

The relational structure $\slott_n$, for $n\in \mathbb N$, is defined as follows.

It has two vertices, $b_n$ and $c_n$. $A(b_n)$, $T(b_n,b_n)$ and $T(c_n,c_n)$ hold in $\slott_n$.
And, for each binary relation symbol $R\in {\alfab}$, there is:

$$\slott_n \models R(b_n,b_n), R(b_n,c_n), R(c_n,c_n)$$

Notice that $\slott_n \not\models T(b_n,c_n)$ and  $\slott_n \not\models T(b_n,c_n)$. This will be important
for Observation \ref{homojest6} to hold.

Obviously, for each $m,n\in \mathbb N$, structures $\slott_n$ and $\slott_m$ are isomorphic, and we will
use the notation $\slott$ when discussing their isomorphism invariant properties (with $b$ and $c$ being
the two verticies of $\slott)$.

We think of structures $\slott_n$ as {\em slots}, into which conjunctive queries of interest can be inserted. Indeed,
 it is easy to verify that:

\begin{observation}\label{homojest1}
For each $1\leq k \leq \kkk$ there exists a homomorphism from
$\jest{\gamma_k} $
to $\slott$. Each such homomorphism maps the distinguished variable of $\jest{\gamma_k} $ to $b$.
\end{observation}

\begin{observation}\label{homojest2}
For each $1\leq k \leq \kkk$ there exists a homomorphism from
 $\jest{\beta_{[l,k]}} $
to $\slott$.

Also, for each $1\leq k \leq \kkk$ there exists a homomorphism from
 $\jest{\beta_{[r,k]}} $
to $\slott$.
\end{observation}

\begin{observation}\label{homojest4}
There exists a homomorphism from
$\jest{\gamma_{\footnotesize \diamondsuit }}$
to $\slott$. Each such homomorphism maps the distinguished variable of $\jest{\gamma_{\footnotesize \diamondsuit }}$ to $b$.
\end{observation}

~

For $\nnn\in \mathbb N$ by $\DDD_{\nnn}$ we will mean the structure being the disjoint union of $\DDD$ and of slots $\slott_1$, $\slott_1$, $\ldots \slott_{\nnn}$.

A close cousin of $\slott_n$ is the ontology $\Omega_n$. In short, $\Omega_n$ postulates the existence of $\slott_n$. Formally speaking,
$\Omega_n$ is a conjunction of the statements $A(b_n)$,  $T(b_n,b_n)$, $T(c_n,c_n)$  and of $R(b_n,b_n), R(b_n,c_n), R(b_n,c_n)$ for each $R\in {\alfab}$.

Finally, for $\nnn\in\mathbb N$ we define:

$$\Omega^{\nnn}\;\; = \;\; \bigwedge_{1\leq n \leq \nnn} \Omega_n $$

\begin{lemma}\label{lem:slott-n-spelnia}
$\DDD_{\nnn} \models  \mathcal O$ and $\DDD_{\nnn} \models \Omega_{\nnn} $.
\end{lemma}

\noindent
{\sc Proof:} Notice that $\slott$ was constructed with conditions (p1) and (p2) from Section \ref{sec:candidate} in mind.\qed


\section{Proof of Theorem \ref{main1}}\label{secmain1}

Now we have all the tools ready to prove Theorem \ref{main1}.

Well, almost all, we need one more definition.
For each $R\in \alfab$ define $\gamma_R(x_R)$ as the conjunctive query with one free variable:
$$ \gamma_R(x_R) \;\; = \;\; \exists y \;\; R(x_R,y)\wedge A(y) $$
Like in Section \ref{sec:slots}, we can easily notice that:

\begin{observation}\label{homojest5}
For each $R\in \alfab$ there exists a homomorphism from
$\jest{\gamma_R} $
to $\slott$. Each such homomorphism maps the distinguished variable of $\jest{\gamma_R} $ to $b$.
\end{observation}

Now we can finally define our Boolean conjunctive query $\psi$ and the ontology \ontology. We will
begin from $\psi$. Let:
$$ {\mathcal I}\;\;=\;\;\alfab \cup \{ \diamondsuit \} \cup \{1,2,\ldots \kkk\}. $$
 $\mathcal{I}$ is meant to be a set of subscripts, and we use it to define query $\psi$ which is defined as the existential closure of:
$$ \bigwedge_{i\in \mathcal I} \gamma_i(x_i)\;\;\;\; \wedge \;\;\;\; \bigwedge_{i,j\in \mathcal I, \;i\neq j} x_i\neq x_{j} $$

\vspace{2mm}
Recall that the UCQ $\Psi$ from the proof of Theorem \ref{niemain1}, in Section \ref{sec:detecting1}, was a disjunction of
 $\jest{\gamma_{\footnotesize \diamondsuit }}$ and of all the queries $\jest{\gamma_k}$ for $1\leq k \leq \kkk$. Now, the idea is
that instead of their disjunction, we take their conjunction, and we additionally require that the distinguished variables
of each two components are interpreted as two distinct vertices of the structure. This almost would work (if accompanied by a right ontology), but only almost:
 we also need (as the Reader is going to see in the proof of Lemma \ref{lem:main1-pos}), one more category of components in this conjunction, namely all the conjunctive queries of the form $ \jest{\gamma_R}$,
 and also for them  we require that all their distinguished variables are interpreted as distinct vertices.

Now let us construct the ontology \ontology. Recall that $\mmm = |\alfab|$ and put $\nnn=\kkk+\mmm$.
Then:
$$ \text{ \ontology} \;\;\;=\;\; {\mathcal O} \;\; \wedge \;\; \Omega^{\nnn} $$
where ${\mathcal O}$ is as defined in Section \ref{sec:candidate} and $\Omega^{\nnn}$ as defined in
Section \ref{sec:slots}.

Now, when \ontology \; and $\psi$ are defined, what remains to be proven are two lemmas:

\begin{lemma}\label{lem:main1-neg}
If  $\thue$ is a negative instance then \ontology  $\;\not\models \psi$.
\end{lemma}

\begin{lemma}\label{lem:main1-pos}
If $\thue$ is a positive instance then \ontology  $\;\models \psi$.
\end{lemma}

\noindent
{\sc Proof of Lemma \ref{lem:main1-neg}}:\vspace{0.1mm}

\noindent
Assume that $\thue$ is negative.
We need to show that a structure $D$ exists, such that $D\models\;$\ontology ~but $D\not\models \psi$.
This $D$ is (as all our Readers have already guessed) the structure $\DDD_{\nnn} $. We already know,
from Lemma \ref{lem:slott-n-spelnia}, that $\DDD_{\nnn}  \models\;$\ontology. What remains to be shown is that
$\DDD_{\nnn} \not\models \psi$.

So assume, towards contradiction, that  $\DDD_{\nnn} \models \psi$.

As we explained in Section \ref{sec:remark}, this means that
there exists a set ${\mathcal H}=\{h_i: i\in {\mathcal I}\} $ of homomorphisms,
with $h_i: \jest{\gamma_i}\rightarrow \DDD_{\nnn}$ for each $i\in \mathcal I $, and  such that if $i\neq j$ then
$h_i$ and $h_j$ map the distinguished variables of $\jest{\gamma_i}$ and $\jest{\gamma_j}$ to two different vertices of
$\DDD_{\nnn}$.

Clearly, the image of each $h\in \mathcal H$ is a connected substructure of $\DDD$.
So let us recall what are the connected substructures of $\DDD_{\nnn}$:

-- there is $\DDD$

-- and there are $\nnn$ ``slots'', i.e. disjoint copies of $\slott$.

We know from Observations \ref{homojest1}, \ref{homojest4} and \ref{homojest5} that for each
$i\in \mathcal I$ there exists a homomorphism from  $\jest{\gamma_i}$ to $\slott $.
But we also know from the same observations that such homomorphism maps the distinguished variable of  $\jest{\gamma_i}$ to $b $. Which means, that
 for every slot $\slott_n$, for $ 1\leq  n \leq \nnn$ there may be at most one $i$ such that the image
 of $h_i$ is in $\slott_n$.

 And recall that we have $\nnn$ slots at our disposal, and there are $\nnn+1$ homomorphisms in $\mathcal H$.

 So there must be at least one $i\in \mathcal I$ such that $h_i$ is a homomorphism from $\jest{\gamma_i}$
 to $\DDD$.

 But $\DDD$ is perfect, so (by Observation \ref{obs-jeden}) no such homomorphism can exist for any $i\in \{1,2,\ldots \kkk \}$.

 And, since we assumed that $\thue$ is negative, we know (by Lemma \ref{lem-dwa}) that there is no homomorphism
 from $\jest{\gamma_\diamondsuit}$ to $\DDD$.

Notice vertex $a$ is the only element of relation $A$ in $\DDD$. But $\DDD \not\models R(x,a)$  for any $R\in \alfab$.
So there is no homomorphism from $\jest{\gamma_R}$ to $\DDD$.

Which contradicts the assumption that
 $\DDD_{\nnn} \models \psi$.
\qed

\vspace{3mm}

\noindent
{\sc Proof of Lemma \ref{lem:main1-pos}:}\vspace{0.5mm}

\noindent
Assume  that $\thue$ is positive, and let $D$ be any structure, with the set $V$ of verticies,
such that $D\models$ \ontology. We need to show that
$D\models \psi$. Since $D\models\;$\ontology,\; we know that
constants $a$, and $b_n, c_n,$ for each $1\leq n\leq \nnn$, are in $V$.
Thanks to the Unique Name Assumption we know that
the constants are interpreted as parwise distinct elements of $V$.
Denote by ${\mathcal B}\subseteq V$ the set
of all the (interpretations of) constants  $b_n$ for each $1\leq n\leq \nnn$.

\begin{lemma} \label{lem:upychanie}
There exist such $\jjj\in \mathcal I$ and such homomorphism $h:\jest{\gamma_\jjj}\rightarrow D$,  that
$h(x_\jjj)\not\in \mathcal B$, where $x_\jjj$ is, as usual, the distinguished variable of $\jest{\gamma_\jjj}$.
\end{lemma}

Before we prove Lemma \ref{lem:upychanie}, let us explain how it will imply
Lemma \ref{lem:main1-pos}.

In order to satisfy $\psi$, we need to provide one homomorphism
$h_i: \jest{\gamma_i}\rightarrow D$ for each $i\in \mathcal I $,
and this must be done in such a way that the distinguished variables
of distinct queries are mapped to distinct vertices of $D$.

Since $\psi$ has  $\nnn+1$ components, we need to provide $\nnn+1$ such homomorphisms.

Having the $h$ from Lemma \ref{lem:upychanie} in hand, we have $\nnn$ queries left to satisfy.
And, when satisfying them, the respective homomorphisms cannot map any distinguished variable to $h(x_\jjj)$ any more,
because this vertex is already occupied by the distinguished variable of $\jest{\gamma_\jjj}$.

But we have $\nnn$ disjoint copies of $\slott$ in $D$ for our disposal,
and we know, again
 from Observations \ref{homojest1}, \ref{homojest4}, and \ref{homojest5}, that each such copy is sufficient to satisfy any
$\jest{\gamma_i}$. And, we also know from these observations, that
when $\slott_n$ is used to satisfy some $\jest{\gamma_i}$ then the distinguished variable of $\jest{\gamma_i}$
 is mapped to $b_i$. Which is not equal to $h(x_\jjj)$, since, thanks to Lemma \ref{lem:upychanie}, we know that
 $h(x_\jjj)\not\in\mathcal B$.

\vspace{2mm}
\noindent
{\sc Proof of Lemma \ref{lem:upychanie}}:

\noindent
Call $D$ {\em dirty} if that there exists a nonempty word $w\in \alfab^*$ and  $b\in \mathcal B$, such that $\strz{a}{w}{b}$.
And call $D$ {\em tidy} otherwise.

We need to consider {\bf three cases}.

\vspace{1.3mm}
\noindent
{\sc Case 1.} $D$ is dirty.
\vspace{0.5mm}

\noindent
If this is the case, let $w=vR$ be a shortest  word such that $\strz{a}{w}{b}$ for some  $b\in \mathcal B$
(here $v$ may be empty).
Then there exist $s\in V$ which is the penultimate vertex on this path from $a$ to $b$, which means that $D\models R(s,b)$ for some
$R\in \alfab$ and $s\not\in \mathcal B$.

Then take $\jjj$ equal to $R$, which means that:
$$\jest{\gamma_{\jjj}}\;\;\;\; \text{is}\;\;\;\; \exists x_R \exists y \;\; R(x_R,y)\wedge A(y) $$
Define $h(x_R)=s$ and $h(y)=b$. This is a homomorphism, because \ontology \; requires that $D\models A(b)$.

\vspace{1.3mm}
\noindent
{\sc Case 2.}  $D$ is tidy and perfect.
\vspace{0.5mm}

\noindent
Then, by Observation \ref{lem-cztery},
 there exists  vertex $s$ of $D$ such that
$\strz{a}{l}{s}$ and $\strz{a}{r}{s}$. We take $\jjj= \diamondsuit$ and $h(x_\jjj)=a$.

\vspace{1.3mm}
\noindent
{\sc Case 3.}  $D$ is tidy and imperfect.
\vspace{0.5mm}

\noindent
In this case, by Observation \ref{obs:brakuje1}, there exists $w\in \alfab^*$,
 $s\in V$,
 and $1\leq k \leq \kkk$,
 such that $\strz{a}{w}{s} $ and
 $D\models  \gamma_{k}(s)   $. Take $\jjj= k$ and $h(x_\jjj)=s$.
We assumed that $D$ is tidy, so we know that $s\not\in \mathcal B$.

This ends the proof of Lemma \ref{lem:upychanie}, Lemma \ref{lem:main1-pos} and of Theorem \ref{main1} \qed


\section{Proof of Theorem \ref{main2}}\label{secmain2}

Now we have all the tools one needs to prove Theorem \ref{main2}.
Almost all. We just need one more definition.

For each $R\in \alfab$ define
 $\beta_R(x_R)$ and $\beta_{\bar{R}}(x_{\bar{R}})$ as  conjunctive queries with safe negation, with one free variable each:
$$ \beta_R(x_R) \;\; = \;\; \exists y,z \;\; T(x_R,y)\wedge R(y,z) \wedge \neg T(x_R,z) $$
$$ \beta_{\bar{R}}(x_{\bar{R}}) \;\; = \;\; \exists y,z \;\; T(x_{\bar{R}},y)\wedge R(z,y) \wedge \neg T(x_{\bar{R}},z) $$

Now, like in Section \ref{sec:slots} we can notice that:

\begin{observation}\label{homojest6}
For each $R\in \alfab$ there exists a homomorphism from
$\jest{\beta_R} $
to $\slott$ and there exists a homomorphism from
$\jest{\beta_{\bar{R}}} $
to $\slott$ .
\end{observation}

At this point we can  define our Boolean conjunctive query $\phi$ and the ontology \ontologyy. We will
begin from $\phi$. Let:
$$ {\mathfrak I}\;\;=\;\;\alfab \cup \{\bar{R}: R\in \alfab \}  \cup   \{ \diamondsuit \} \cup \{[l,k],[r,k]: 1 \leq k \leq  \kkk\} $$
Query $\phi$ is now defined as the existential closure of:
$$ \bigwedge_{i\in {\mathfrak I}} \beta_i(x_i)\;\;\;\; \wedge \;\;\;\;
\bigwedge_{i,j\in {\mathfrak I}, \;i\neq j,\;R\in \alfab}
 \neg R(x_i,x_{j})  $$

We act here in the same spirit as we did  in Section \ref{secmain1}. We replace the disjunction of conjunctive queries, from Section
\ref{sec:detecting2}, with their conjunction (including however some additional components) and
  we additionally require that the distinguished variables
of each two components are interpreted as  vertices of the structure not connected by any binary relation of
$\bar{\alfab} $. We can do it, because we are  allowed to use safe negation in $\phi$. Recall that in Section \ref{secmain1} we had inequality at our disposal, so we required the distinguished variables to be interpreted as distinct vertices.

Now let us construct the ontology \ontologyy. Recall that $\mmm = |\alfab|$ and put $\nnn=2(\kkk+\mmm)$.
Then:
$$ \text{ \ontologyy} \;\;\;=\;\; {\mathcal O} \;\; \wedge \;\; \Omega^{\nnn} $$
where ${\mathcal O}$ is as defined in Section \ref{sec:candidate} and $\Omega^{\nnn}$ as defined in
Section \ref{sec:slots}.

Now, once \ontologyy \; and $\phi$ are defined,  two lemmas  remain to be proven:

\begin{lemma}\label{lem:main2-neg}
If  $\thue$ is a negative instance then \ontologyy  $\;\not\models \phi$.
\end{lemma}

\begin{lemma}\label{lem:main2-pos}
If $\thue$ is a positive instance then \ontologyy  $\;\models \phi$.
\end{lemma}

\noindent
{\sc Proof of Lemma \ref{lem:main2-neg}}:\vspace{0.5mm}

\noindent
Assume that $\thue$ is negative.
We need to show that a structure $D$ exists, such that $D\models\;$\ontologyy ~but $D\not\models \phi$.
This $D$ is (as all the Readers have already guessed for the second time) the structure $\DDD_{\nnn} $. We already know,
from Lemma \ref{lem:slott-n-spelnia}, that $\DDD_{\nnn}  \models\;$\ontologyy. What remains to be shown is that
$\DDD_{\nnn} \not\models \phi$.

So assume, towards contradiction, that  $\DDD_{\nnn} \models \phi$.

As we explained in Section \ref{sec:remark}, this means that
there exists a set ${\mathfrak H}=\{h_i: i\in {\mathfrak I}\} $ of homomorphisms,
with $h_i: \jest{\beta_i}\rightarrow \DDD_{\nnn}$ for each $i\in \mathfrak I $, and  such that if $i\neq j$ then
$h_i$ and $h_j$ map the distinguished variables of $\beta_i$ and $\beta_j$ to  vertices that are not connected in
$\DDD_{\nnn}$ by any relation from $\bar{\alfab}$.

Clearly, the image of each $h\in \mathfrak H$ is a connected substructure of $\DDD_{\nnn}$.
So let us recall what are the connected substructures of $\DDD_{\nnn}$:

-- there is $\DDD$

-- and there are $\nnn$ ``slots'', i.e. disjoint copies of $\slott$.

We know from Observations \ref{homojest2}, \ref{homojest4} and \ref{homojest6} that for each
$i\in \mathcal I$ there exists a homomorphism from  $\jest{\beta_i}$ to $\slott $.

But the (two) vertices of each slot are connected by relations from $\alfab$, and each of them is also connected to itself, via loops.
For this reason,
for every slot $\slott_n$, for $ 1\leq  n \leq \nnn$ there may be at most one $i$ such that the image
 of $h_i$ is in $\slott_n$.

 And again, like in Section \ref{secmain1},  we have $\nnn$ slots at our disposal, and there are $\nnn+1$ homomorphisms in $\mathfrak H$.

 So there must be at least one $i\in \mathfrak I$ such that $h_i$ is a homomorphism from $\jest{\beta_i}$
 to $\DDD$.

 But $\DDD$ is perfect, so (by Observation \ref{obs:nienazwana}) no such homomorphism can exist for any
  $i\in \{[l,k],[r,k]: 1 \leq k \leq  \kkk\}$.

 And, since we assumed that $\thue$ is negative, we know (by Corollary \ref{lem-cztery-cor}) that there is no homomorphism
 from $\jest{\beta_\diamondsuit}$ to $\DDD$.

Notice also that $a$ is the only vertex satisfying $\DDD \models \exists y \;\; T(x_R,y) $, but
on the other hand $\DDD \models \; T(a,s) $ holds for each vertex $s$ of $\DDD$. So
neither $\jest{\beta_R} $ nor $\jest{\beta_{\bar{R}}}  $ can be satisfied in $\DDD$ for any $R\in \alfab$.

 Which contradicts the assumption that
 $\DDD_{\nnn} \models \phi$.
\qed

\vspace{3mm}
\noindent
{\sc Proof of Lemma \ref{lem:main2-pos}}:\vspace{1mm}

\noindent
Assume  that $\thue$ is positive, and let $D$ be any structure, with the set $V$ of verticies,
such that $D\models$ \ontologyy. We need to show that
$D\models \phi$.

Since $D\models\;$\ontologyy\; we know that
constants $a$, and $b_n, c_n,$ for each $1\leq n\leq \nnn$, are in $V$.

Denote by ${\mathfrak B}$ the set of all the (interpretations in $D$) of constants $b_n$ and $c_n$ for
$1\leq n \leq \nnn$.

Thanks to the Partial Closed Word Assumption we know that there are no relational atoms in $D$, connecting two
vertices of $\mathfrak B$, or connecting an element of $\mathfrak B$ with $a$, apart from the facts specified by \ontologyy.
In particular, there is no atom in $D$ connecting $a$ and any element of $\mathfrak B$, and, if $n\neq m$, then there is no atom
 in $D$ connecting some $b_n$ or $c_n$ with $b_m$ or $c_m$.

Call $D$ {\em soiled} if there exist $s\in V$, $R\in \alfab$, $w\in \alfab^*$  and $d\in \mathfrak B$  such that
$\strz{a}{w}{s} $ and $R(s,d)$, or $R(d,s)$, hold in $D$. And call $D$ {\em clean} otherwise. The
crucial property of soiled structures is:

\begin{observation}\label{lem:soiled}
If $D$ is soiled then there exist vertices $s,t$ of $D$, and $R\in \alfab$ such that $D\models T(a,s)\wedge R(s,t) \wedge \neg T(a,t)  $
or such that $D\models T(a,s)\wedge R(t,s) \wedge \neg T(a,t)$.
\end{observation}

The following lemma is the counterpart of Lemma \ref{lem:upychanie}:

\begin{lemma} \label{lem:upychanie2}
There exist such $\jjj\in \mathfrak I$ and such homomorphism $h:\jest{\beta_\jjj}\rightarrow D$,  that
neither $S( h(x_\jjj), d)$ nor $S(d, h(x_\jjj))$ holds in $D$, for any $S\in\bar{\alfab}$,
and any $d\in \mathfrak B$,
where $x_\jjj$ is, as usual, the distinguished variable of $\jest{\beta_\jjj}$.
\end{lemma}

Before we prove Lemma \ref{lem:upychanie2}, let us explain how it will imply
Lemma \ref{lem:main2-pos}.

Like in the proof of Lemma \ref{lem:main1-pos}, in order to satisfy $\phi$, we need to provide one homomorphism
$h_i: \jest{\beta_i}\rightarrow D$ for each $i\in \mathfrak I $,
but this time this must be done in such a way that the distinguished variables
of distinct queries are mapped to unrelated vertices of $D$.

Since $\psi$ has  $\nnn+1$ components, we need to provide $\nnn+1$ such homomorphisms.

Having the $h$ from Lemma \ref{lem:upychanie2} in hand, we have $\nnn$ queries left to satisfy.
And, when satisfying them, the respective homomorphisms cannot map any distinguished variable to any
vertex which is connected, by any atom of any relation from $\bar{\alfab}$, with $h(x_\jjj)$.

But we have $\nnn$ disjoint copies of $\slott$ in $D$ for our disposal,
and we know,
 from Observations \ref{homojest2},  \ref{homojest4}, and    \ref{homojest6},  that each such copy is sufficient to satisfy  any
$\jest{\beta_i}$. And, we also know from these observations, that
when $\slott_n$ is used to satisfy some $\jest{\beta_i}$ then the distinguished variable of $\jest{\beta_i}$
 is mapped to $b_n$ or $c_n$. Which is --  thanks to Lemma \ref{lem:upychanie2}  -- not related to $h(x_\jjj)$.

\vspace{2mm}
\noindent
{\sc Proof of Lemma \ref{lem:upychanie2}}:
Like in the proof of Lemma  \ref{lem:main1-pos}, there are {\bf three cases}:

\vspace{1.3mm}
\noindent
{\sc Case 1.} $D$ is soiled.

\noindent
Then, by Observation \ref{lem:soiled}
 there exist vertices $s,t$ of $D$, and $R\in \alfab$, such that $D\models T(a,s)\wedge R(s,t) \wedge \neg T(a,t)  $
or such that $D\models T(a,s)\wedge R(t,s) \wedge \neg T(a,t)  $.

Take $h(x_\jjj)=a$ and $\jjj = R$ or $\jjj=\bar{R}$, depending on whether
$D\models  R(s,t)$ or  $D\models  R(t,s)$.

 By Partial Closed Word Assumption, neither $S(a,d)$ nor $S(d,a)$
holds in $D$ for any $S\in \alfab$ and any $d\in \mathfrak B$

\vspace{1.3mm}
\noindent
{\sc Case 2.}   $D$ is clean and perfect.

\noindent
Then, like in the proof of Lemma \ref{lem:upychanie}, we use Observation \ref{lem-cztery},
 which gives us  vertex $s$ of $D$ such that
$\strz{a}{l}{s}$ and $\strz{a}{r}{s}$. We take $\jjj= \diamondsuit$ and $h(x_\jjj)=a$.
Again, by Partial Closed Word Assumption, neither $S(a,d)$ nor $S(d,a)$
holds in $D$ for any $S\in \alfab$ and any $d\in \mathfrak B$

\vspace{1.3mm}
\noindent
{\sc Case 3.}  $D$ is clean but imperfect.

\noindent
 Then, by Observation
 \ref{obs:brakuje2},
 there exist
 $w\in \alfab^*$,
 $s\in V$,
 and $1\leq k \leq \kkk$,
 such that $\strz{a}{w}{s} $ and
 $D\models  \beta_{[l,k]}(s) \vee  \beta_{[r,k]}(s)  $.

 We take $h(x_\jjj)=s$ and $\jjj= [l,k]$ or   $\jjj= [r,k]$, depending on whether
  $D\models  \beta_{[l,k]}(s)$ or $D\models \beta_{[r,k]}(s)  $.

Since we assume that $D$ is clean, we can be sure that  neither $S(s,d)$ nor $S(d,s)$
holds in $D$ for any $S\in \alfab$ and any $d\in \mathfrak B$.

\section{Remark about finite entailment} \label{sec:finite}

 Instead of query entailment, sometimes finite query entailment is considered. In this scenario,
 condition:

 \noindent
 {\em for every structure $D$, being a model of $\kontologia$ , query $\Xi$ is true in $D$}

 from
 Section \ref{nuance}, is
 replaced with:

 \noindent
{\em for every finite structure $D$, being a model of $\kontologia$ , query $\Xi$ is true in $D$.}

Our results, and their proofs, extend, almost without any changes, to the finite entailment scenario, and let us explain why.

It is well known, that the following two sets:

\noindent
(${\mathfrak S}_1$)~~ Set of  triples $[l,r, \Pi]$ such that $l\szymeq r$

\noindent
(${\mathfrak S}_2$)~~  Set of  triples $[l,r, \Pi]$, such that there exists a finite semigroup, in which the equality $l=r$ does not
hold, but all the equalities $l_k=r_k$ for $1\leq k\leq \kkk$ do hold.

are recursively inseparable.

Now, to prove undecidability results, analogous to Theorems \ref{main1} -- \ref{niemain2}, but for finite
entailment, it is enough to construct, for given $[l,r, \Pi]$, an ontology $\kontologia$  and a conjunctuve query $\xi$
(with inequalities, or with safe negation), such that:

\noindent
\textbullet~ if $[l,r, \Pi]\in {\mathfrak S}_1$, then for each structure $D$ such that  $D\models \kontologia $ there is also $D\models \xi $;

\noindent
\textbullet~ if $[l,r, \Pi]\in  {\mathfrak S}_2$, then there is a finite structure $D$ such that  $D\models \kontologia $ but $D\not\models \xi $.

It is easy to see however, that the ontologies and queries we constructed in Sections \ref{secmain1} and \ref{secmain2}
satisfy the above requirements. For the second implication, just replace the $\DDD$ from our construction with the
finite semigroup whose existence is guaranteed by the assumption.


\appendix

\bibliographystyle{abbrv}
\bibliography{__main}

\begin{thebibliography}{1}

\bibitem{handbook}
J.~Barwise, editor.
\newblock {\em Handbook of Mathematical Logic}.
\newblock North-Holland, New York, 1977.

\end{thebibliography}

\end{document}